\documentstyle[multicol,psfig,aps]{revtex}
\begin{document}
\draft
\title{EXCITED STATES IN $^{52}$Fe AND THE ORIGIN OF THE YRAST TRAP AT
I$^\pi$=12$^+$}
%insert title here
\author{C.A. Ur$^{1,2}$, D. Bucurescu$^2$, S.M. Lenzi$^1$, 
G. Mart\'\i nez-Pinedo$^3$,
D.R. Napoli$^4$, \\
D. Bazzacco$^1$, F. Brandolini$^1$, D.M. Brink$^5$, J.A. Cameron$^6$, 
G.de Angelis$^4$, M. De Poli$^4$, \\
A. Gadea$^4$, S. Lunardi$^1$, N. M\u arginean$^2$, M.A. Nagarajan$^7$, 
P. Pavan$^1$, \\
C. Rossi Alvarez$^1$, C.E. Svensson$^6$}
\address
{$^1$Dipartimento di Fisica and INFN, Sezione di Padova, Padova, Italy}
\address
{$^2$H.Hulubei National Institute of Physics and Nuclear Engineering, Bucharest,
Romania}
\address
{$^3$W.K. Kellogg Radiation Laboratory, California Institute of Technology,
Pasadena, USA}
\address
{$^4$Laboratori Nazionali di Legnaro, INFN, Legnaro, Italy}
\address
{$^5$Dipartimento di Fisica and INFN, Trento, Italy}
\address
{$^6$McMaster University, Ontario, Canada}
\address
{$^7$Department of Physics, UMIST, Manchester, United Kingdom}
\date{\today}
\maketitle
\begin{abstract}
  Excited states in $^{52}$Fe have been determined up to spin
  10$\hbar$ in the reaction $^{28}$Si + $^{28}$Si at 115 MeV by using
  $\gamma$--ray spectroscopy methods at the GASP array. The excitation
  energy of the yrast 10$^+$ state has been determined to be 7.381
  MeV, almost 0.5 MeV above the well known $\beta^+$--decaying yrast
  12$^+$ state, definitely confirming the nature of its isomeric
  character. The mean lifetimes of the states have been measured by
  using the Doppler Shift Attenuation method. The experimental data
  are compared with spherical shell model calculations in the full
  $pf$-shell.
\end{abstract}

\begin{center}
\end{center}
                                

% body of paper here
\begin{multicols}{2}
  
\section{Introduction}

Recently, the study of the $f_{7/2}$--shell nuclei has gained renewed
interest. The development of very efficient detector arrays both for
$\gamma$--rays and charged particles has allowed investigation of the
structure of these nuclei at high spins. Close to the middle of the
$f_{7/2}$ shell, nuclei show strong collective behaviour near the
ground state~\cite{cam94,len96,len97}.  At higher spins, shape
transitions towards triaxial and non-collective deformations can occur
due to an intimate interplay between the collective and microscopic
degrees of freedom. Recently, band terminating states, corresponding
to fully aligned $f_{7/2}$ configurations, were observed in
$^{48}$Cr~\cite{len96} and $^{50}$Cr~\cite{len97}.  When approaching
the doubly magic nucleus $^{56}$Ni the collective behaviour is rapidly
disappearing as nuclei evolve towards a spherical shape. So far, the
nucleus $^{52}$Fe has been an experimental challenge. Most of the
known excited states in this nucleus are of relatively low spin (below
6$\hbar$) and have been observed in
($^3$He,n)~\cite{alf75,boh75,iri77}, ($\alpha$,2n)~\cite{eve77} and
(p,t)~\cite{dec78} reactions (see also ref.~\cite{huo94}).  Attempts
to extend its yrast structure to higher spins in fusion-evaporation
reactions induced by heavy ions have failed so far (see
refs.~\cite{avr76,gee76} and references therein). The reason of this
has been attributed to the probable inversion between the lowest
10$^+$ and 12$^+$ states, which results in an isomeric 12$^+$ yrast
trap that decays by ${\beta}^+$ emission to
$^{52}$Mn~\cite{gee76,GMM75}. Thus, although fusion-evaporation
reactions populate appreciably the $^{52}$Fe channel, the deexcitation
$\gamma$--ray flux effectively stops at the 12$^+$ level where it is
diverted into the population of high spin states in $^{52}$Mn. Due to
the weak direct population of the lower spin states, it was not
possible to observe the yrast line above the 4$^+_1$
state~\cite{gee76}. One should note, however, that these previous
experiments were performed with small (10-15$\%$ efficiency) Ge(Li)
detectors, and therefore had a rather low $\gamma$--ray detection
efficiency.

In this work, performed with a powerful $\gamma$--ray detector array,
we have been able to extend the level scheme of $^{52}$Fe up to the
10$^+$ state, thereby confirming the predicted inversion of the 10$^+$
and 12$^+$ states.

The experimental data have been interpreted in the framework of
spherical shell model (SM) calculations in the full {\it pf} shell.

\section{Experimental details}

High spin states in the nucleus $^{52}$Fe have been populated via the
$^{28}$Si + $^{28}$Si reaction at 115 MeV beam energy.  The silicon
beam was delivered by the XTU Tandem accelerator at the National
Laboratory of Legnaro. The target was a 0.8 mg/cm$^2$ $^{28}$Si foil
evaporated on a 13 mg/cm$^2$ Au backing. Gamma rays have been detected
with the GASP array~\cite{baz92} which consists of 40
Compton-suppressed large volume HP Ge detectors and the 80 BGO
crystals inner ball. The 40 Ge detectors are placed symmetrically
relative to the beam axis on seven rings as follows: 6 detectors at
35$^\circ$, 6 detectors at 60$^\circ$, 4 detectors at 72$^\circ$, 8
detectors at 90$^\circ$, 4 detectors at 108$^\circ$, 6 detectors at
120$^\circ$ and 6 detectors at 145$^\circ$.  Data have been recorded
when at least two Ge detectors and two elements of the BGO inner ball
fired in coincidence. We collected on tape a total of 7.8 x 10$^8$ two
fold and 5.4 x 10$^7$ three fold events.  Gain matching and efficiency
calibration of the Ge detectors have been performed using $^{152}$Eu
and $^{56}$Co radioactive sources.  The total cross section of the
reaction used was fragmented in a large number of reaction channels.
In order to estimate the relative yield of these channels we have used
a total $\gamma$--$\gamma$ coincidence matrix in which we determined
the intensities of the $\gamma$--ray transitions feeding the ground
states.  The most intense channels populated in our reaction were
$^{50}$Cr ($\alpha 2p$) with $\sim$31\%, $^{49}$Cr ($\alpha 2pn$) with
$\sim$29\% and $^{49}$V ($\alpha 3p$) with $\sim$11\% of the total
cross section. States in $^{52}$Fe have been populated either by
$2p2n$ or $\alpha$ evaporation. The population of the ground state of
$^{52}$Fe was estimated to represent less than 1\% of the total gamma
yield.

\section{Data analysis}
\subsection{The level scheme}

Previous studies of $^{52}$Fe~\cite{iri77,dec78,gee76,via71} have
established the position of the yrast 2$^+$, 4$^+$ , 3$^-$ and 5$^-$
states, of a second 4$^+$ state and of the long lived 12$^+$ state.
This isomeric level, placed at an excitation energy of 6820$\pm$130
keV and with a measured half-life of 45.94$\pm$1.0 sec~\cite{gee76},
decays $\beta^+$ towards states in $^{52}$Mn. An upper limit of 0.4 \%
was established~\cite{gee76} for the $\gamma$--decay of the 12$^+$
state.  The high efficiency of the GASP array and the use of a
reaction in which low-lying levels are considerably populated by side
feeding allowed us to identify new $\gamma$--ray transitions belonging
to $^{52}$Fe by setting gates on the previously known $\gamma$--rays.
Double gated $\gamma$--coincidence spectra with gates set on some key
transitions assigned to $^{52}$Fe are shown in Fig.~1. On the basis of
such $\gamma$--$\gamma$ coincidence data obtained from a
$\gamma$--$\gamma$--$\gamma$ coincidence cube and of the relative
intensities of the transitions, we have constructed the level scheme
shown in Fig.~2.  The relative intensities of the transitions have
been extracted from the 90$^\circ$ spectrum in coincidence with the
850 keV 2$^+$$\rightarrow$0$^+$ transition in order to avoid the
uncertainties introduced by the lineshape broadening.  The intensity
for the 2753 keV transition which is assumed to be of $\Delta$I=1
character (see level scheme) is already corrected for the angular
distribution as specified in ref.~\cite{yam67}. The high energy part
of the spectrum at 90$^\circ$ in coincidence with the 850 keV
transition, from which the relative intensities were extracted, is
shown in Fig.~3.

We could confirm the yrast 6$^+$ state suggested in ref.~\cite{iri77};
furthermore we established a new 6$^+$ state as well as two 8$^+$
states and a 10$^+$ state lying at 7.381 MeV, well above the 12$^+$
$\beta^+$ decaying isomer. This constitutes the first experimental
evidence for the predicted~\cite{gee76} inversion between the first
10$^+$ and 12$^+$ states.  Three new $\gamma$--rays connecting the
5$^-$ level to the 3$^-$, 4$^+_1$ and 4$^+_2$ states have been
identified.  We have to mention here that high spin states in
$^{52}$Fe have been also identified in a parallel experiment performed
recently at Gammasphere~\cite{rie97}.

The low statistics together with the high background produced 
by the more strongly populated nuclei in this reaction, prevented the 
observation 
of a possible E4 $\gamma$--ray connecting the 12$^+$ and the 8$^+$ states.

Spins and parities have been assigned on the basis of the angular 
distribution of the $\gamma$--rays. Data were sorted 
in two $\gamma$--$\gamma$ coincidence matrices having on one axis 
$\gamma$--rays detected at all angles and on the second axis those detected 
at 60$^\circ$ and 120$^\circ$ and those detected at 90$^\circ$, respectively. 
By setting gates on the axis with all the detectors, the intensity of
the observed $\gamma$--rays follows the regular angular distribution law
disregarding the multipole character of the gating transition.
The ADO (Angular Distribution from Oriented states) ratio is defined as
\cite{pii96}:
$$R_{ADO} = {{(I_\gamma(\theta)+I_\gamma(\pi 
-\theta))/2}\over{I_\gamma(90^\circ)}}$$

\noindent
where $I_\gamma$ denotes the intensity of the observed $\gamma$--ray at the 
angles
$\theta$, $\pi -\theta$ and 90$^\circ$, respectively, corrected by the 
detection efficiency. Typical values of the ADO ratios for
$\theta$=60$^\circ$ in the GASP geometry are $\sim$ 1.17 for a stretched
quadrupole transition and $\sim$ 0.85 for a stretched dipole transition.
The $\gamma$--ray energies and relative intensities of the transitions
belonging to $^{52}$Fe, together with their ADO ratios at 60$^\circ$ and 
spin--parity
assignments are reported in Table \ref{table1}. Spin and parity of the new 
levels are
based on ADO analysis, by assuming that transitions with
R$_{ADO}\approx$ 1.17 have stretched E2 character. The results for the 1941, 
2168 and 889
keV transitions lead therefore to I$^\pi$ = 10$^+$ for the state
at 7.381 MeV.   

We could not extract the angular distribution of the 2735 keV and 2753 keV
$\gamma$--ray transitions depopulating the 4$_2^+$ and 5$^-$ states, 
respectively, since their broadened lineshapes are overlapping.
The spin assignment for the states decaying via these two transitions 
are based on previous measurements. In ref.~\cite{iri77} the
angular distribution of the 2735 keV transition was found to be compatible
with an E2 character. This assignment is now supported by the observation of a
1286 keV transition connecting  
the 6$^+_2$ state to the 4$^+_2$ state. 
A level at 5138$\pm$4 keV
excitation energy  decaying towards the 4$^+$ yrast state was also reported in 
ref.~\cite{iri77}; later on~\cite{dec78},
a 5$^-$ assignment has been given to that state.
This brings a $\Delta$I=1 character for the new measured
$\gamma$--ray transitions of 1553 keV and 2753 kev.
The 5$^-$ assignment is also confirmed by the E2 character of the 740 keV
$\gamma$-ray feeding 
the 3$^-$ state.

\subsection{DSAM analysis}

To perform the analysis of the Doppler broadened lineshapes we sorted
the data in seven 4k x 4k $\gamma$--$\gamma$--coincidence matrices, each
corresponding to the coincidence between the detectors of one ring and all 
other detectors.

We have analyzed the $\gamma$--ray spectra in coincidence with the 850 keV 
$\gamma$--ray (which does not show any appreciable broadening) 
in order to select 
better the channel of interest and to reduce as much as possible
the contaminations on the lineshapes of the relevant transitions.

The 10$^+$$\rightarrow$ 8$_1^+$ and 10$^+$$\rightarrow$ 8$_2^+$ 
$\gamma$--ray transitions do not exhibit a 
broadened lineshape, 
indicating that the 10$^+$ state has a long lifetime (longer than a few ps). 
Changes in the lineshape for states below were observed only when a large 
amount of side feeding was present (see Table \ref{table1}). 

The lifetime analysis was carried out with the computer code 
LINESHAPE~\cite{wel91}. The slowing down process and the scattering of the 
recoils in 
the target and in the backing were described by a Monte Carlo simulation as it 
was 
developed by Bacelar {\it et al.}~\cite{bac87}, with a modification regarding 
the spread in the initial direction of the recoils due to the evaporation of 
light particles~\cite{bra97}. The simulation was performed with 5000 
histories and up to 187 time steps covering the recoil range in 
the backing. 
Northcliffe and Schilling~\cite{NS70} electronic stopping power values have
been used in the calculations.

The 
program performs a $\chi^2$ minimization of the lineshape fit as a function of 
the level lifetime, the side-feeding time and the normalization factors.
We have used a one step side-feeding for each level, 
the side-feeding intensity being a fix parameter in the program.  
The background and the intensity of contaminant peaks present in the spectra 
have been kept fixed. The analysis was 
done for each line separately starting with the highest transition 
in the level scheme. 
Lineshapes were fitted at forward and backward angles simultaneously allowing 
a better identification of the contaminants. 
The program, designed to deal with cascades of $\gamma$--rays connecting states
with side-feeding originating only from the continuum, is not suited 
for level schemes such as that of $^{52}$Fe. Anyway, it could be used
also in our case by properly transforming the complex feeding scheme in 
equivalent $\gamma$--ray cascades.
We have first determined the lifetimes of the two 8$^+$ states
from the analysis of the two decay branches, 
10$^+\rightarrow$ 8$^+_1\rightarrow$ 6$^+_1$ and
10$^+\rightarrow$ 8$^+_2\rightarrow$ 6$^+_1$, respectively.
The lifetime of the yrast 6$^+$ state has been extracted by analyzing the decay
pattern along the yrast sequence. To account for the feeding provided via the
decay of the 8$_2^+$ state, we have introduced above the 6$^+_1$ state a
virtual level with zero 
lifetime and side-feeding time given by the lifetime of the 8$^+_2$ state. The
intensity of the 1021 keV transition has been modified to account for the whole
longlived population of the 6$^+_1$ level provided by the decay of the 10$^+$ 
state. 
The lifetime of
the second 6$^+$ state has been extracted from the analysis of the
sequence of $\gamma$--ray transitions 
10$^+\rightarrow$ 8$^+_2\rightarrow$ 6$^+_2\rightarrow$ 4$^+_1$.
In order to describe the multiple feeding of the 4$^+_1$ state,
we have again introduced virtual levels of mean lifetime equal to zero and  
side-feeding times given by the lifetimes of the 6$^+_2$ and 5$^-$ states,
respectively. In Fig.~4 the best fits for the lineshapes (measured at 
72$^\circ$) of several $\gamma$--ray transitions are displayed. 
The lineshape of the 2035 keV transition is strongly contaminated
by the presence of the 2045 keV line belonging to $^{49}$Cr with a very
pronounced lineshape. We have determined the lifetime of the 8$^+_1$ state 
from the best fit of the experimental spectrum with that obtained after summing
the 
calculated lineshape of the 2035 keV $\gamma$--ray transition and the
experimental lineshape of the contaminant line.
The two lines 
of 2735 and 2753 keV overlapped their lineshapes and consequently, we could 
not extract a definite lifetime value 
for the 4$_2^+$ and 5$^-$ states. We estimated a lower limit of about 1 ps
for the sidefeeding of the 4$_1^+$ provided via the 2753 keV transition. 
We have varied this sidefeeding time in the range 0.5--2.0 ps without 
practically affecting the 
mean lifetime value of the 4$_1^+$ state. The obtained results are reported 
in Table \ref{table2} together with the previous measured values. The 
experimental reduced transition probabilities $B(E2)$ have been extracted 
according to the expression~\cite{eji89}:
$$B(E2)\; (e^2b^2)={{0.08156\; B_\gamma}\over{\tau\; E_\gamma^5\; 
(1+\alpha_{tot})}}$$

\noindent
where $B_\gamma$ is the branching ratio of the $\gamma$--ray transition, 
$\tau$ is the lifetime of the state in picoseconds, $E_\gamma$ is the energy of 
the transition in MeV and $\alpha_{tot}$ is the total conversion coefficient.  
These $B(E2)$ values are compared to the ones calculated within 
the SM (see Sect. IV.A).

\section{DISCUSSION}
\subsection{Shell Model Calculations}

The structure of $^{52}$Fe has been analyzed in the
framework of the spherical shell model in the full $pf$ shell ($m$-scheme
dimension 109,954,620).  These
calculations have shown to reproduce with very good accuracy the
experimental data of the nuclei in this mass 
region~\cite{len96,len97,cau94,cau95,mar97}.  
The single particle
energies were taken from the $^{41}$Ca experimental spectrum and the
effective interaction used was the KB3~\cite{PZ81}. The
effects of core polarization on the quadrupole properties were taken
into account by using the effective charges $q_{\pi}$=1.5 and
$q_{\nu}$=0.5. The hamiltonian was treated by the Lanczos method and
diagonalized with the code ANTOINE~\cite{antoine}.
The resulting theoretical level scheme is compared in Fig.~5 with the
experimental one. A fairly good agreement is found. The energy inversion
of the 12$^+$ isomeric state with the 10$^+$ yrast is reproduced
theoretically, even if the energy gap between the two states is
smaller than in the experiment.

A reasonable reproduction of the energy spectrum is not enough to establish the 
goodness of the shell model calculations. A more stringent test is their
ability to reproduce the experimental lifetimes or the $B(E2)$ values.  
The calculated $B(E2)$ values are displayed in Table~\ref{table2}. They are in 
good agreement with the experimental ones. The values for the 6$^+_2$ and 
the 8$^+_2$ states seem 
to be not very well described by the calculations but this can be due to the
vicinity in energy to the states of the same spin and parity (see Fig.~2). 

The shell model calculations provide also the spectroscopic quadrupole 
moments $Q_{spec}$ which are plotted in the lowest panel of Fig.~6.
Large negative values are obtained for the first two excited states. 
As already pointed out in
ref.~\cite{MG83}, $^{52}$Fe behaves as a rotor below $I=6\hbar$, consistently 
with a $K=0$ band. 
Using the rotational model prescription we obtained for the 2$^+$ and 4$^+$
states an intrinsic quadrupole moment $Q_0 \approx 90~ $e~fm$^2$ from both 
the theoretical $B(E2)$ values and the spectroscopic quadrupole moments. The 
deduced
deformation parameter is $\beta=0.23$. At $I=6\hbar$, $Q_{spec}$ changes 
sign and takes a very small value. This change of regime can be associated to
the process of particle alignment.
In this mass region, the most deformed nuclei lay in the middle of the shell.
In particular, the nucleus $^{48}$Cr has the maximum number
of particles to develop quadrupole collectivity.  At high spin, the
interplay between single particle and collective degrees of freedom
produces changes of shape towards spherical or non collective oblate
states.  This is consistent with the fact that the mechanism of
generating angular momentum by aligning the valence particle spins
along the rotational axis in a high-j shell becomes energetically
favoured at high frequency. Nuclei as $^{52}$Fe which do not lay near the 
middle of the shell, are not much deformed and the incipient rotational 
behaviour at low spin smears out very soon with increasing angular momentum.

Recently, it has been shown that the development of
quadrupole coherence that gives rise to rotational--like bands 
in these nuclei~\cite{zu48}, is originated by the  
mixing of the $f_{7/2}$ and $p_{3/2}$ orbits. 
It is thus interesting to follow the single orbital contributions to the 
wavefunction as the angular momentum increases.
In Fig.~6, the fractional occupation numbers of the $f_{7/2}$ and
$p_{3/2}$ orbits are plotted for the yrast states of $^{52}$Fe. For comparison, 
the same quantities are also reported for its cross conjugate
nucleus $^{44}$Ti and for the most quadrupole deformed nucleus in the $f_{7/2}$ 
shell, $^{48}$Cr. 
In the simple $(f_{7/2})^n$ model, $^{52}$Fe and $^{44}$Ti should 
have the same energy spectra (cross conjugate symmetry). 
In fact, the two level schemes are similar at low spin. However, the symmetry
is lost at high spin: there is no inversion of the 10$^+$ and 12$^+$ states
in $^{44}$Ti. Then, even if from a 
qualitative analysis of Fig.~6 the $f_{7/2}$ subshell results by large the 
most occupied one, the contribution arising from the rest of the orbitals 
in the $pf$ shell becomes crucial to obtain a good description of 
these nuclei. The fractional occupation number of the $p_{3/2}$ orbital is 
much more 
important than the other two orbitals $f_{5/2}$ and $p_{1/2}$ (not included in 
the figure).   At low spin, the $p_{3/2}$ occupation number 
remains almost constant for all three nuclei. The biggest $p_{3/2}$
contribution is observed for the most deformed nucleus $^{48}$Cr, where it 
begins to decrease at the backbending ($I=10\hbar$). In 
$^{52}$Fe this happens much earlier, at $I=6\hbar$. 

 At the maximum spin 
that can be constructed with the valence particles in the $f_{7/2}$ shell, the
$p_{3/2}$ occupation number vanishes for $^{44}$Ti ($I=12\hbar$) and
becomes insignificant for $^{48}$Cr ($I=16\hbar$).  Also the smaller 
contribution from the
$f_{5/2}$ and $p_{1/2}$ decreases at high spin and at the band
terminating states the $f_{7/2}$ becomes the only relevant orbit.
These fully-aligned band-terminating states are of non collective character.  
The situation is different in $^{52}$Fe, where above $I=6\hbar$, the
$p_{3/2}$--shell contribution stays almost constant as a function of
spin and the same happens for the other components, even at
$I=12\hbar$.  There is still some quadrupole coherence in the 12$^+$ state. 
This can be related to the energy inversion of the 10$^+$ and 12$^+$ states, 
which gives rise to the yrast trap, as will be shown below. 

The change of regime at $I=6\hbar$, reflected in the fractional 
occupation numbers and in the spectroscopic quadrupole moments, 
can be related in the rotational 
model to a crossing between the ground state $K=0$ band and an excited 
$K=6$ band. To study this problem we have
computed different Nilsson intrinsic states and projected them onto good 
angular momentum. The $K=0$ band corresponds to an intrinsic state obtained 
by filling the [330]1/2, [321]3/2, [312]5/2 Nilsson orbitals for protons and
neutrons. The intrinsic state of the $K=6$ band is constructed by exciting one
proton {\it or} one neutron from the [312]5/2 to the [303]7/2 orbital.
Our calculations indicate that the $I=6$ $K=6$ and
the $I=6$ $K=0$ are degenerate. This also explains the presence of two
6$^+$ levels close in energy (see Fig.~2). The states between the 6$^+$ 
and the 10$^+$ could be thus considered as a mixing of a $K=0$ and a $K=6$ 
band. We cannot speak of a well defined intrinsic state.
 
We repeated this procedure for the state at $I=12\hbar$, where there is a
residual quadrupole coherence. A $K=12$ intrinsic prolate Nilsson state can be 
constructed by exciting two 
particles, a proton {\it and} a neutron, from the [312]5/2 to the 
[303]7/2 orbit. After projecting this 
state onto good angular momentum, an overlap of $\sim$~0.9 with the exact shell 
model wave function of the 12$^+$ state is obtained. The excitation energy of 
this level is lower than the $I=10\hbar$ and $I=12\hbar$ states coming from the 
$K=0$ and $K=6$ bands, which explains the presence of the yrast trap.

\subsection{The systematics of yrast traps}

Excited nuclear states in the vicinity of closed shells are
explained through the rearrangement of the nucleons in the available
single particle orbitals. This results usually in irregular decay
patterns and isomeric states, often called, at high spins,
``yrast traps''. The isomeric 12$^+$ state in $^{52}$Fe represents an 
energy spin yrast trap due to the large difference of its spin with 
those of the lower states.

Several high spin yrast traps have been identified along the periodic
table but only few of them originate through the inversion in energy
with yrast states of lower spins.  Such yrast traps have been
identified in $^{53}$Fe~\cite{GMM75}, $^{53}$Co~\cite{esk66},
$^{156}$Hf~\cite{sev96}, $^{211}$Po~\cite{man60} and $^{212}$Po~\cite{per62}. 
The nature of the energy spin traps can 
be related~\cite{ring} to the alignment of nucleons. The nucleons with
aligned spins gain energy because their residual interaction is stronger
for wave functions with large spatial overlap (MONA
effect~\cite{fae76}). In most of the cases the structure of these states is
associated with oblate deformation or ``single particle'' rotations.
A different structure is, however, found in $^{52}$Fe where the effect of the
maximum alignment is accompanied by a quadrupole coherence that gives rise to a
prolate deformation at $I=12\hbar$, as stated in the previous subsection. 

In Fig.~7a, the yrast bands of the even-even N=Z nuclei $^{44}$Ti, $^{48}$Cr 
and $^{52}$Fe are shown up to the maximum spin that can be built in 
the $f_{7/2}$ shell. The nucleus $^{48}$Cr,
is presented to stress the evolution of the collectivity along the N=Z
line. The similarity between the level structure for $^{44}$Ti and
$^{52}$Fe can be seen up to spin 10$\hbar$ but in $^{44}$Ti, even if
the energy of the 12$^+$ state is lowered due to the alignment of the valence
nucleons, no inversion is observed.
As discussed in the previous subsection, in $^{44}$Ti and $^{48}$Cr
the $I=12\hbar$ and $I=16\hbar$ levels, respectively, behave as non
collective band terminating states with pure $f_{7/2}$ configurations.
On the other hand, the
12$^+$ state in $^{52}$Fe still preserves contributions from the
$p_{3/2}$ orbital, the nucleus is slightly prolate deformed and therefore more
collective.

In Fig.~7b, one can see the systematics of other yrast traps in this
mass region.  The nucleus $^{53}$Fe and its mirror $^{53}$Co present
yrast traps at $I=19/2\hbar$ at an excitation energy around 3 MeV that
lay below the 15/2$^-$ states.  The 19/2$^-$ level in $^{53}$Fe decays
by $\gamma$ emission of E4, M5 and E6 character, while in $^{53}$Co decays 
$\beta^+$. In these cases the yrast traps are also well reproduced by
the shell model calculations. In the Nilsson calculation, we obtain a 
prolate $K=19/2$ state for $^{53}$Fe ($^{53}$Co) by exciting one of the protons
(neutrons) from the [312]5/2 to the [303]7/2 and coupling them to the odd
neutron (proton).   
As in the case of the $N=Z$ $^{52}$Fe, these aligned collective states are
favoured in energy and become yrast traps. These 
mirror nuclei do not behave as their cross conjugates $A=43$. In fact,
in $^{43}$Ti and $^{43}$Sc there are isomeric 19/2$^-$ states at about
3~MeV excitation energy but higher in energy with respect to the 15/2$^-$ 
states to which they decay by $\gamma$ emission.  The systematics shows that 
the cross conjugate symmetry is not verified in these nuclei.  
One can interpret these differences in terms of
deformation or available valence space. In fact,  
at the beginning of the shell, $^{44}$Ti together with the mirrors
$A=43$, have a reduced number of particles and the fully aligned
states are of pure $(f_{7/2})^n$ configuration of non-collective character.  
For the heavier group
of nuclei, the maximum spin state is slightly prolate deformed due to the
contribution of the $p_{3/2}$ orbit and yrast traps can be formed.

\section{CONCLUSIONS}

We have investigated high spin states in $^{52}$Fe in a backed target
experiment. The high detection efficiency of the GASP array and the
relatively large amount of side-feeding population of the states of interest 
allowed us to establish the level scheme up to spin
$I=10\hbar$, a fact which was hindered until now by the presence of the
12$^+$ yrast trap. We could therefore establish experimentally that
the retardation of the decay of the 12$^+$ yrast state is due to its
location at an energy lower than that of the 10$^+$ state.  The mean
lifetime of the states have been determined by means of the Doppler
Shift Attenuation Method.

The experimental results have been compared with calculations
performed within the spherical SM in the {\it pf} shell. Reasonably
good agreement has been found for both the energy of the states and
the $B(E2)$ values. The difference between the 12$^+$ states in the
cross conjugate nuclei $^{44}$Ti and $^{52}$Fe can be satisfactorily
explained through shell model calculations. It was concluded that
apart from the lowering of the states due the maximal overlapping of
the nucleonic wave functions, in $^{52}$Fe this state is even lower
due to a higher degree of collectivity present in the structure of the
wave function related to the contribution of the $p_{3/2}$ orbital.

\acknowledgements

The authors would like to thank B.F. Bayman, E. Caurier, A. Poves, D.
Rudolph and A.P. Zuker for fruitful and interesting discussions.  We
also thank the crew of the XTU Tandem of the National Laboratory of
Legnaro for the smooth operation of the accelerator during the
experiment. The computational cycles were provided by the Centro de
computaci\'on cient\'{\i}fica de la Facultad de Ciencias at the
Universidad Aut\'onoma de Madrid. G.M.P. was partly supported by the
DGICyES (Spain) and A.G. was supported by the EC under contract number
ERBCHBGCT940713.

% now the references. delete or change fake bibitem. delete next three
%   lines and directly read in your .bbl file if you use bibtex.

\newpage
\narrowtext

\begin{figure}[h]
  \begin{center}
    \leavevmode
    \psfig{file=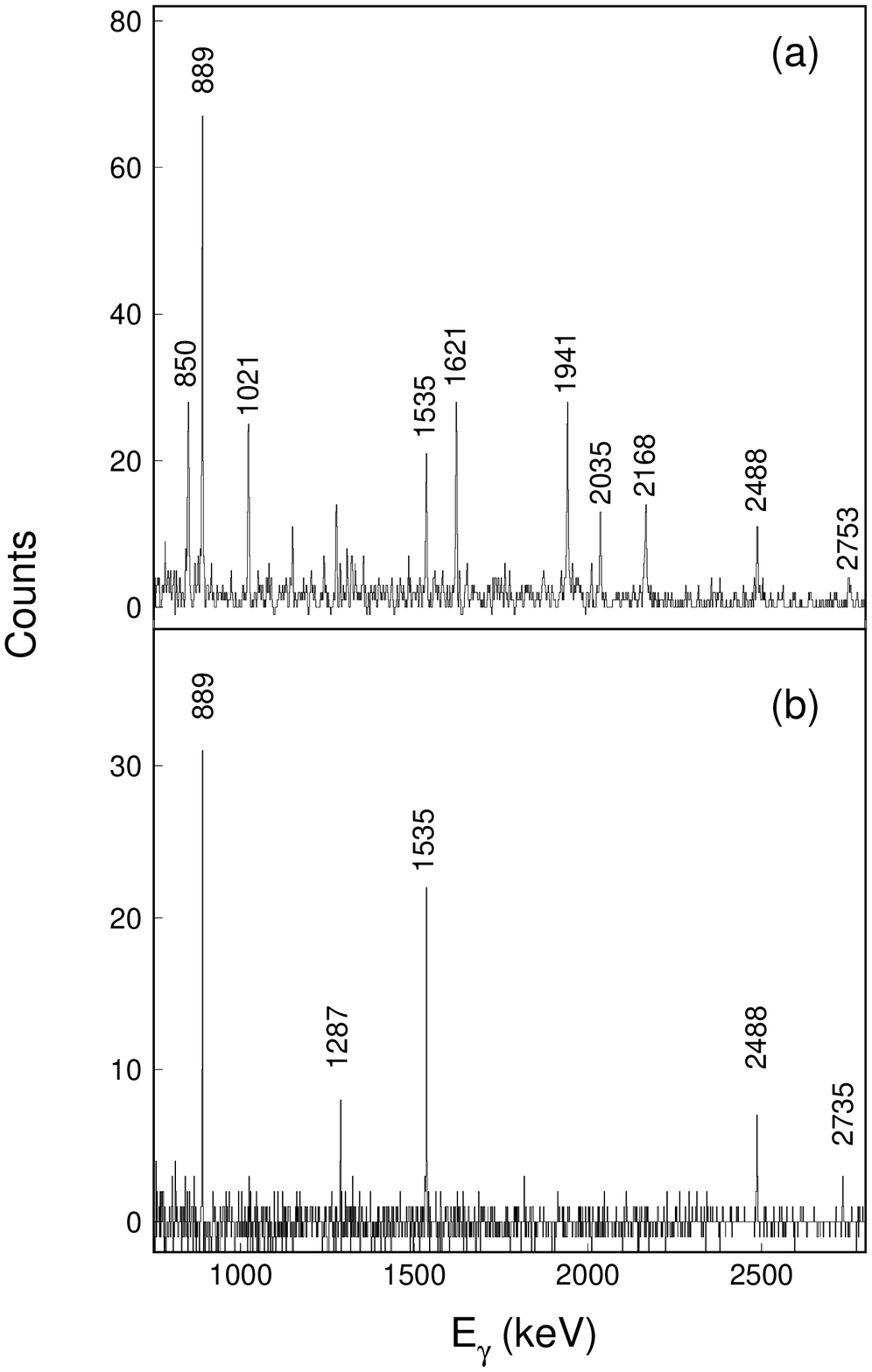,width=0.45\textwidth}
    \caption{Examples of double-gated spectra for some selected transitions
      assigned to $^{52}$Fe: a) gates set on the first two yrast
      transitions, 2$^+_1$$\rightarrow$0$^+_1$ and
      4$^+_1$$\rightarrow$2$^+_1$; the contaminant peaks between 1021
      keV and 1535 keV belong to the strong channel $^{47}$V; b) gates
      set on the 2$^+_1$$\rightarrow$0$^+_1$ and
      8$^+_2$$\rightarrow$6$^+_2$ transitions.}
    \label{fig:fig1}
  \end{center}
\end{figure}

\begin{figure}[h]
  \begin{center}
    \leavevmode
    \psfig{file=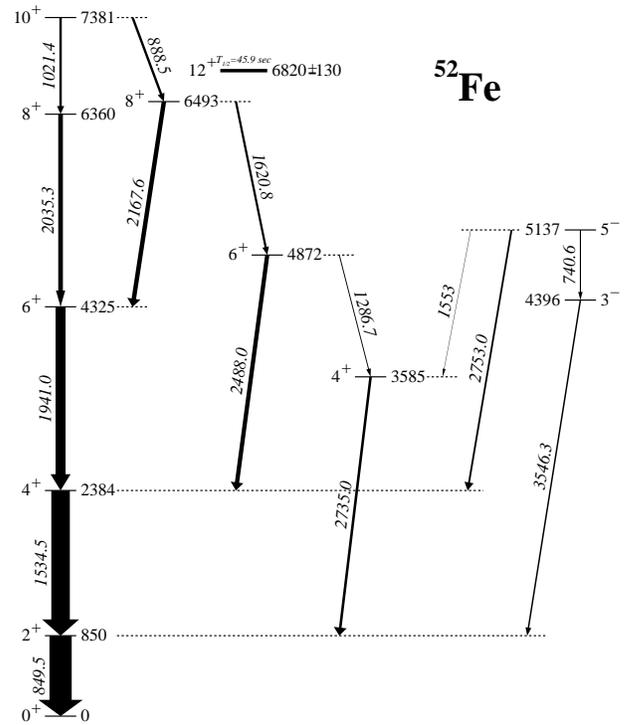,width=0.45\textwidth}
    \caption{Level scheme of $^{52}$Fe, as obtained
      in the present experiment. The excitation energy of the 12$^+$
      $\beta^+$-decaying isomeric state is taken from
      ref.~\protect\cite{gee76}}
    \label{fig:fig2}
  \end{center}
\end{figure}

\begin{figure}[h]
  \begin{center}
    \leavevmode
    \psfig{file=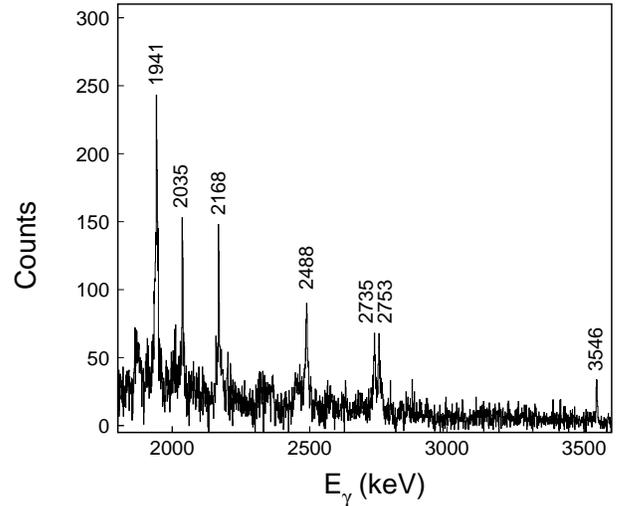,width=0.45\textwidth}
    \caption{High energy $\gamma$--ray spectrum at 90$^\circ$ in 
      coincidence with the 850 keV transition.}
    \label{fig:fig3}
  \end{center}
\end{figure}

\begin{figure}[th]
  \begin{center}
    \leavevmode
    \psfig{file=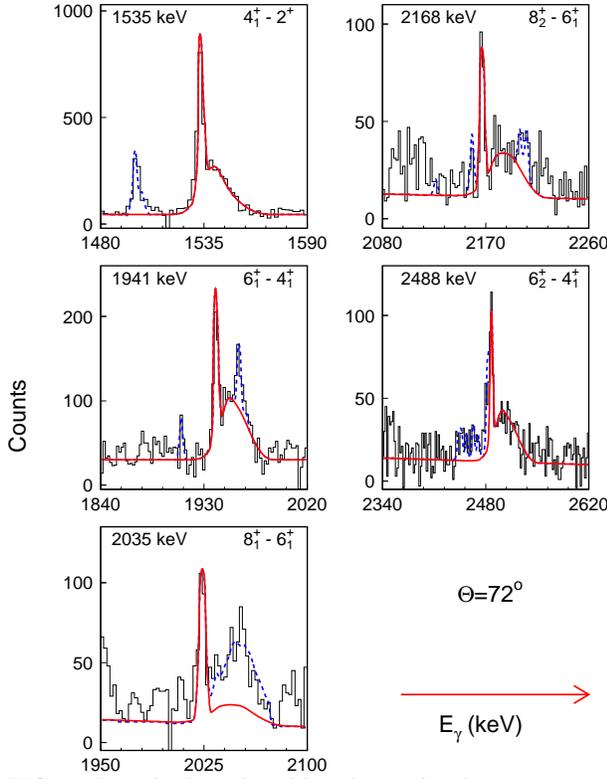,width=0.45\textwidth}
    \caption{Doppler broadened lineshapes for the transitions of 1535 keV
      (4$^+_1$$\rightarrow$2$^+_1$), 1941 keV
      (6$_1^+$$\rightarrow$4$^+_1$), 2035 keV
      (8$_1^+$$\rightarrow$6$^+_1$), 2168 keV
      (8$^+_2$$\rightarrow$6$_1^+$) and 2488 keV
      (6$_2^+$$\rightarrow$4$^+_1$), at 72 degrees. Full lines show
      least square fits performed with the LINESHAPE program and
      correspond to the mean lifetime values reported in
      Table~\protect\ref{table2}. The contaminant peaks are indicated
      by dashed lines.}
    \label{fig:fig4}
  \end{center}
\end{figure}

\begin{figure}[bh]
  \begin{center}
    \leavevmode
    \psfig{file=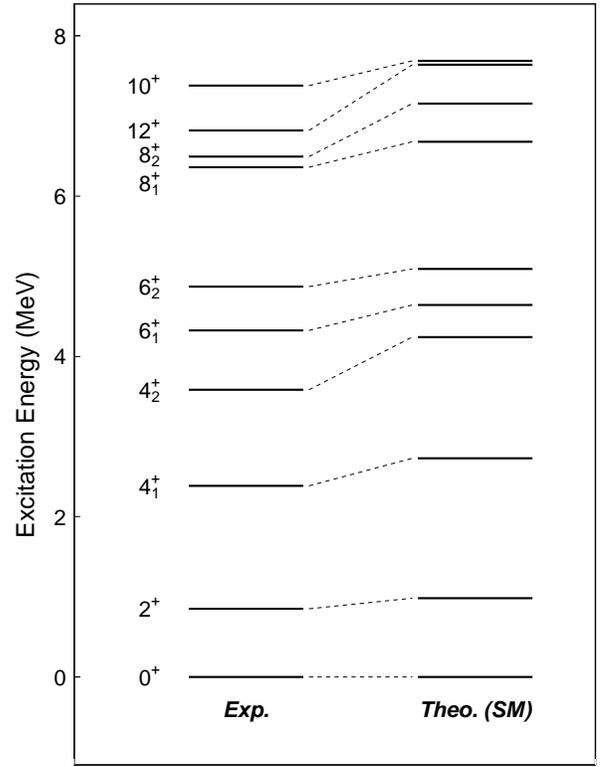,width=0.45\textwidth}
    \caption{Comparison between the experimental and the shell model
      positive parity energy levels in $^{52}$Fe. Dashed lines connect
      levels of the same spin parity. The calculations reproduce the
      inversion between the 10$^+$ and 12$^+$ states.}
    \label{fig:fig5}
  \end{center}
\end{figure}

\end{multicols}

\widetext

\begin{table}
\caption{Relative intensities, ADO ratios and spin assignments in $^{52}$Fe.}
\label{table1}
\begin{tabular}{cccc}
E$_\gamma$ (keV) & Intensity\tablenotemark[1]\tablenotemark[2]& ADO
ratios\tablenotemark[1]  &  
Assignment \\
\hline
~740.6        & ~~~~~~~5.5(~6)& 1.27(11) & $~5^-\rightarrow ~3^-$     \\
~849.5        &        -      &    -     & $~2^+\rightarrow ~0^+$     \\
~888.5        & ~~~~~~11.5(~8)& 1.20(~8) & $10^+\rightarrow ~8^+_2$   \\
1021.4        & ~~~~~~13.1(25)&    -     & $10^+\rightarrow ~8^+_1$   \\    
1286.7        & ~~~~~~~5.0(10)&    -     & $~6^+_2\rightarrow ~4^+_2$ \\
1534.5        & ~~~~100.0(~6) & 1.16(~4) & $~4^+_1\rightarrow ~2^+$   \\
1553~~        & ~~~~~~~1.0(~5)&    -     & $~5^-\rightarrow ~4^+_2$   \\
1620.8        & ~~~~~~13.6(26)&    -     & $~8^+_2\rightarrow ~6^+_2$ \\
1941.0        & ~~~~~~55.0(30)& 1.15(~6) & $~6^+_1\rightarrow ~4^+_1$ \\
2035.3        & ~~~~~~21.0(30)& 1.46(18) & $~8^+_1\rightarrow ~6^+_1$ \\
2167.6        & ~~~~~~20.7(20)& 1.24(11) & $~8^+_2\rightarrow ~6^+_1$ \\
2488.0        & ~~~~~~21.9(15)& 1.34(19) & $~6^+_2\rightarrow ~4^+_1$ \\
2735.0        & ~~~~~~15.0(17)&    -     & $~4^+_2\rightarrow ~2^+$   \\
2753.0        & ~~~~~~10.0(20)&    -     & $~5^-\rightarrow ~4^+_1$   \\
3546.3        & ~~~~~~~7.0(15)& 0.92(~8) & $~3^-\rightarrow ~2^+$     \\
\end{tabular}
\tablenotetext[1]{Values are extracted from spectra in coincidence with the
850 keV 2$^+$$\rightarrow$0$^+$ transition.}
\tablenotetext[2]{Errors higher than 10\% are specified.}
\end{table}

\begin{multicols}{2}
\narrowtext

\begin{minipage}[h]{0.45\textwidth}
\begin{figure}[h]
  \begin{center}
    \leavevmode
    \psfig{file=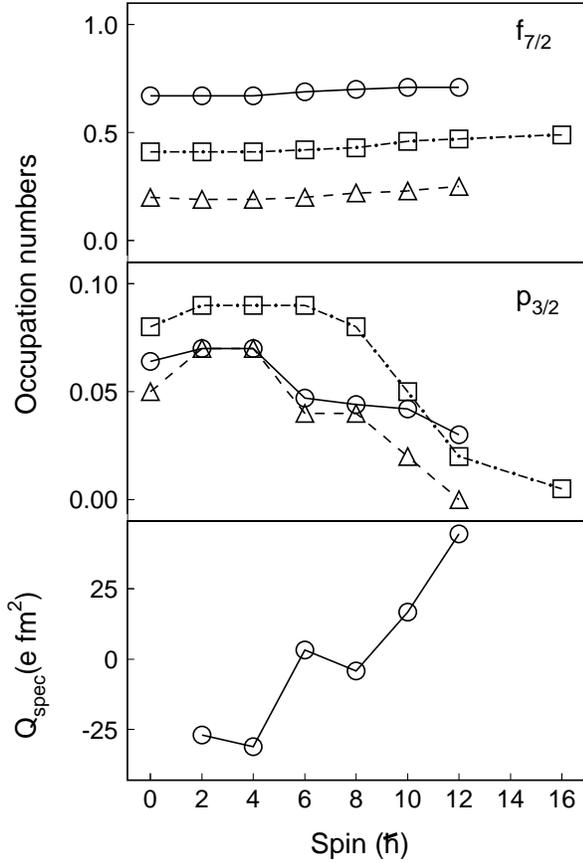,width=\textwidth}
    \caption{The upper two panels show the occupation numbers of specific 
      orbitals in the {\it pf} shell for the yrast even spin states in
      $^{44}$Ti (triangles), $^{48}$Cr (squares) and $^{52}$Fe
      (circles) as extracted from the SM calculations.  The lowest
      panel show the spectroscopic quadrupole moments of the yrast
      states in $^{52}$Fe obtained with full $fp$ spherical shell
      model calculations.}
    \label{fig:fig6}
  \end{center}
\end{figure}
\end{minipage}

\begin{figure}[h]
  \begin{center}
    \leavevmode
    \psfig{file=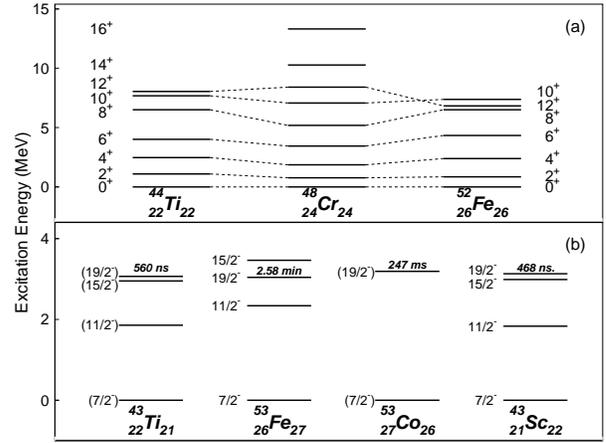,angle=270,width=0.45\textwidth}
    \caption{Systematics of the yrast traps in the $f_{7/2}$ shell. 
      a) Yrast states in the cross-conjugated nuclei $^{44}$Ti and
      $^{52}$Fe and in the self-conjugated nucleus $^{48}$Cr showing
      the evolution of the collectivity along the N=Z line in the
      $f_{7/2}$ shell.  b) Other yrast traps showing the broken
      symmetry between cross conjugate nuclei.}
    \label{fig:fig7}
  \end{center}
\end{figure}

\end{multicols}

\widetext

\begin{table}
\caption{Mean lifetimes of the states in $^{52}$Fe as extracted from the
present analysis compared with the ones of previous works and with the SM 
calculations. Statistical errors are specified for the side-feeding 
times.}
\label{table2}
\begin{tabular}{ccccccccc}
E$_x$ (keV) & E$_\gamma$ (keV) & I$_i$ & I$_f$ &
 \multicolumn{2}{c}{$\tau$ (ps)} 
 &
$\tau_{SF}$ (ps) & \multicolumn{2}{c}{B(E2) ($e^2fm^4$)} \\
\cline{5-6}\cline{8-9}
          & & &  & previous work~\cite{gee76} & present work & & Exp. & SM \\
\hline
~850&~849.5&2~    &0~&$>$1.0                &      -      &         -
  &$<$1844
\tablenotemark[1] &154.5\\
2384&1534.5&4$_1$ &2~   &0.40$^{+0.29}_{-0.14}$&
0.32$\pm$0.08&0.06$^{+0.03}_{-0.02}$ & 300$\pm$69 & 223.5 \\
4325&1941.0&6$_1$ &4$_1$&   -                  &
0.24$\pm$0.08&0.04$^{+0.02}_{-0.01}$ & 124$\pm$40 & 117.9 \\
4872&2488.0&6$_2$ &4$_1$&   -                  &
0.30$\pm$0.12&0.04$^{+0.03}_{-0.02}$ & ~29$\pm$14 & ~83.3 \\
6360&2035.3&8$_1$ &6$_1$&   -                  &
0.21$\pm$0.08&0.01$^{+0.01}_{-0.01}$ & ~74$\pm$25 & ~85.6 \\
6493&2167.6&8$_2$ &6$_1$&   -                  &
0.26$\pm$0.06&0.01$^{+0.01}_{-0.00}$ & ~43$\pm$15 & ~11.3 
\end{tabular}
\tablenotetext[1]{Value calculated on the basis of the experimental limit
given in ref. [11].}
\end{table}
\end{document}